# The digital randomness of black–body radiation

**Sándor Varró**
Wigner Research Centre for Physics, Hungarian Academy of Sciences,
Institute for Solid State Physics and Optics, Budapest, Hungary
E-mail: varro.sandor@wigner.mta.hu

**Abstract.** The statistical properties of the fractional part of the random energy of a spectral component of black–body radiation have been analysed in the frame of classical Kolmogorovian probability theory. Besides the integer part of the energy (which satisfies the well-known Planck–Bose distribution) the realizations of its fractional part (related to 'round-off errors') has been represented by binary sequences, like z = 0.001011000010.... It has been shown that the binary variables realized by the 0-s and 1-s at different positions are *independen*t. From the condition of independence the original distribution of the fractional part z can be recovered. If these binary variables have the same distribution, they describe a temperature-independent random energy, whose expectation value is the well-known zero–point energy. Thus, the zero–point fluctuation can be considered as a physical representative of an ideal random number generator.

## 1. Introduction

In the present paper on 'digital randomness' we mean the following: when we express the measurement data acquired in a physical experiment, then the digits of the representing real numbers are always undergoing some random changes from one measurement act to the other. These uncertainties may come from the imperfect control of the boundary conditions, from the always 'finite resolution' of the measuring apparatus, or from the 'loss of precision' in data acquisition, processing, and in displaying these data. Accurate numbers, like sharp and reproducible sequences of the 1 (yes) and 0 (no) digits within some finite time interval, can be obtained if we can answer with certainty the questions 'what is unity?' and 'what is zero?'. Of course, it is an important matter at what accuracy can we decide (draw the level of some physical quantity for distinction) between these 'yes'-es and 'no'-s. The uncertainties can also be caused by the 'objective randomness' of the constituents and transitions of the physical system under study, which cannot be avoided by any means. In the present paper we shall deal with this kind of randomness, which manifest itself in the energy distribution of black–body radiation.

Thermal noise, due to the universal presence of black–body radiation (Planck, 1959) is the main basic source of unavoidable fluctuations and uncertainties in physical measurements. It has already been emphasized by Planck (1908) that even the splitting of the total energy of a moving body in two parts; to the 'translation energy' and to the 'internal energy', is conceptionally impossible, because the 'thermal part' of this latter energy necessarily contains heat radiation, whose energy (and mass) depend on the velocity of the body. Taking the simple numerical example due to Planck (1908); if an ideal gas is heated at constant volume, then the ratio, at which the acquired energy is distributed among the black-body radiation and the mechanical energy of the molecules, is ¼ at 0.001 Hgmm pressure and temperature 2063 Kelvin (~ the melting point of platinium). It is also well-known that in order to have a computer reliably working, the proper cooling of it has to be secured. Otherwise – as is expected merely on physical

grounds – sooner or later the system becomes to much loaded with thermal noise, and finally a 'physical loss of precision' may occur, even if no parts are mechanically or electronically damaged.

Objective indeterminacy has been widely accepted as an inherent property of quantum systems, and the task of the theory is just to quantify this property. In quantum theory the elementary measurement acts are represented by projectors on Hilbert spaces, which form the non–distributive lattice associated to the physical systems. On the other hand, in real experiments, every piece of information on 'genuine quantum objects' like single photons (Oxborrow and Sinclair, 2005) or biphotons (Bogdanov et al., 2007) is gained through macroscopic interfaces, which are necessarily loaded by additional noise. The 'signal–to–noise ratio' define the reliability of any experiment, and it is calculated from relative frequencies, according to classical probability theory (which is based on the 'common sense' distributive Boole algebra of events). In this context we note that recently, several useful concepts and quantities (like for instance the Rényi entropy) introduced in classical probability have received an increasing importance in quantum optics, too (Man'ko and Man'ko, 2011). In the present paper we base all of our considerations on the classical Kolmogorovian probability theory (see e.g. Feller (1970) or Rényi (1970)).

In our earlier studies concerning black–body radiation, on the basis of classical probability theory we have already proved that from the classical chaotic energy distribution – by subtracting the 'dark', fractional part – one can derive the Planck–Bose distribution (Varró, 2007). The boson gas of photons can further be uniquely decomposed into the assembly of 'binary photons', which follow Fermi–Dirac statistics, and they may serve as a natural physical basis for binary representation of integer numbers (Varró 2006a-b). In the present work we discuss the fractional part of the energy, which may also be considered as a 'round–off error' in the relative count number of photons of energy $h\nu$, or an error in measured correlations of photon count numbers (Varró, 2011). It is assumed that the counter has reached the thermal equilibrium during long enough experimental runs. The (random) fractional part $\zeta$ of the energy is represented by binary sequences e.g. z = 0.001011000010.... We analyse the connection between the distribution of $\zeta$ and the distribution of zeros and ones, which are being realizations of binary random variables.

In order to have the present paper possibly self–contained, in Section 2 we summarize our results on the fractional part and the integer part of the energy of a mode of black–body radiation. Section 3 is devoted to the proof of the independence of the binary components. In Section 3 we summarize the results of the present paper.

## 2. The fractional part and the integer part of the energy of a chaotic field component

In the present section we briefly summarize the main steps of the derivation of the Planck–Bose distribution, by subtracting the 'dark', fractional part from the chaotic energy, according to our earlier work (Varró, 2007). In classical physics the black–body radiation, a radiation being in thermal equilibrium in a *Hohlraum* (a cavity with perfectly reflecting walls at absolute temperature $T$) is considered as a chaotic electromagnetic radiation. The average spatial distribution of such a stationary radiation is homogeneous and isotropic and the electric field strength and the magnetic induction of its spectral components have completely random amplitudes which are built up of infinitely many independent infinitesimal contributions. In this description the electric field strength and the magnetic induction of a mode (characterized by its frequency $\nu$, wave vector and polarization) of the thermal radiation are proportional with the random process

$$a_\nu(t) = a_c \cos(2\pi\nu \cdot t) + a_s \sin(2\pi\nu \cdot t) = \sqrt{a_c^2 + a_s^2}\cos(2\pi\nu \cdot t - \vartheta)\,, \qquad \vartheta = \arg(a_c + i a_s)\,, \tag{1}$$

where $a_c$ and $a_s$ are independent random variables. According to the Central Limit Theorem (see e.g. Feller, 1970), these amplitudes have Gaussian *density functions*,

$$P(q \le a_c < q+dq; p \le a_s < p+dp) = \left\{ \frac{1}{a\sqrt{2\pi}} \exp(-q^2/2a^2)dq \right\} \cdot \left\{ \frac{1}{a\sqrt{2\pi}} \exp(-p^2/2a^2)dp \right\}, \tag{2}$$

where the parameter $a$ is related to the spectral energy density $u_\nu = \left\langle a_\nu^2(t)/8\pi \right\rangle = a^2/8\pi = Z_\nu \bar{\varepsilon}$, with $Z_\nu = 8\pi\nu^2/c^3$ being the well–known spectral mode density in vacuum. The average energy of one mode is denoted by $\bar{\varepsilon}$, and $u_\nu d\nu$ gives the energy density of the chaotic radiation in the spectral range $(\nu, \nu + d\nu)$. By introducing the mode energy as a classical random variable, $E = (a_c^2 + a_s^2)/Z_\nu 16\pi$, and the 'action–angle parameters' $\varepsilon = (q^2 + p^2)/Z_\nu 16\pi$ and $\theta = \arg(q + ip)$, we have from (2)

$$\begin{aligned} &P(q \le a_c < q+dq; p \le a_s < p+dp) \\ &= P(\varepsilon \le E < \varepsilon + d\varepsilon) P(\theta \le \vartheta < \theta + d\theta) = [(1/\bar{\varepsilon})\exp(-\varepsilon/\bar{\varepsilon})d\varepsilon](d\theta/2\pi). \end{aligned} \tag{3}$$

Thus, the mode energy, $E$ satisfies an exponential (Boltzmann type) distribution, which stems from the two-dimensional Gauss distribution (2). Henceforth, we shall not display the uniform phase distribution, represented by the factor $(d\theta/2\pi)$ in (3). In our earlier analysis (Varró, 2007) we have introduced two independent energy parameters, $\varepsilon_0$, $\bar{\varepsilon}$, containing two universal constants, and *derived* their physical meaning, too. $\bar{\varepsilon}$ has turned out to be $\bar{\varepsilon} = kT$, where $k = 1.831 \times 10^{-16} erg/K$ is the Boltzmann constant and $T$ is the absolute temperature of the radiation. $\varepsilon_0 = h\nu$ is proportional with Planck's quantum of action, $h = 6.626 \times 10^{-27} erg \cdot \sec$. In terms of the dimensionless energy variable $\eta \equiv E/\varepsilon_0$, equation (3) can be brought to the form

$$P(y \le \eta < y+dy) = f_\eta(y)dy, \quad f_\eta(y) = \beta e^{-\beta y} \quad (0 < y < \infty), \quad \eta \equiv E/\varepsilon_0, \quad \beta \equiv \varepsilon_0/\bar{\varepsilon}. \tag{4}$$

By introducing the Boltzmann entropy $S_\eta = -k\int_0^\infty f_\eta \log f_\eta dy$, and by using the basic relation $\partial S_\eta/\partial E_\eta = 1/T$ of phenomenological thermodynamics, we immediately have $E_\eta \equiv \varepsilon_0 \bar{\eta} = \bar{\varepsilon} = kT$, the law of equipartition of energy. From (4), the distribution function of a mode reads

$$P(\eta < y) \equiv F_\eta(y) = 1 - e^{-\beta y} \quad (0 \le y < \infty), \quad \eta \equiv E/\varepsilon_0, \quad \beta = \varepsilon_0/kT. \tag{5}$$

According to our earlier work (Varró, 2007), the density function and the distribution function of the fractional part $\zeta \equiv \{\eta\} \equiv \eta - [\eta]$ of the mode energy reads

$$P(\zeta < z) = G_\zeta(z) = \frac{1-e^{-\beta z}}{1-e^{-\beta}}, \quad f_\zeta(z) = \frac{\beta e^{-\beta z}}{1-e^{-\beta}} \quad (0 \le z \le 1), \quad \zeta \equiv \{\eta\} \equiv \eta - [\eta], \tag{6}$$

where $[\eta]$ denotes integer part, i.e. the largest integer which is smaller than or equal to $\eta$. The expectation of the fractional part of the energy is

$$E_\zeta \equiv \varepsilon_0 \bar{\zeta} = \varepsilon_0 \int_0^1 f_\eta(z) z dz = \varepsilon_0 \left[ \frac{1}{(\varepsilon_0/kT)} - \frac{1}{\exp(\varepsilon_0/kT) - 1} \right], \quad \lim_{T\to\infty} E_\zeta = \frac{1}{2}\varepsilon_0, \tag{7}$$

which means that we have derived the Planck factor $[\exp(\varepsilon_0/kT)-1]^{-1}$ from the continuous distribution (6). It is remarkable that for very high temperatures the expectation value of the fractional part tends to a temperature-independent constant, as is shown in the second equation of (7).

On the basis of (6) and (4), we have proved that the *integer part* $\xi \equiv [\eta]$ must satisfy the *Planck–Bose distribution*, i.e.

$$P(\xi = n) = \frac{\overline{n}^n}{(1+\overline{n})^{1+n}}, \quad \xi \equiv [\eta], \quad \overline{n} \equiv \frac{1}{e^{\varepsilon_0/kT}-1} = \overline{n}(\varepsilon_0/kT), \quad E_\xi \equiv \varepsilon_0 \overline{\xi} = \varepsilon_0 \overline{n},$$

$$S_\xi[\{P(\xi=n)\}] \equiv -k\sum_{n=0}^{\infty} P(\xi=n) log P(\xi=n) = k[(1+\overline{n})log(1+\overline{n}) - \overline{n} log \overline{n}]. \tag{8}$$

In the last equation of (8) we have also shown the definition and the closed form of the entropy $S_\xi$ of the integer part $\xi$. From Wien's displacement law (1894, see Planck (1959)), written in the form $S_\xi = \Psi(\nu/T)$, where $\Psi$ an universal function, it follows from (8) that the energy parameter $\varepsilon_0$ must be proportional with the frequency, i.e. $\varepsilon_0 = h\nu$. The constant of this proportionality is just Planck's universal quantum of action $h$, and then the universal function is given in (8). According to $\varepsilon_0 = h\nu$ and the relation $\eta = [\eta] + \{\eta\} \equiv \xi + \zeta$, from (8), (7) and (4) , for the spectral energy density $u_\nu$ we obtain the well-known *Planck's law of black–body radiation* (Planck, 1959),

$$u_\nu = \frac{8\pi\nu^2}{c^3}\frac{h\nu}{e^{h\nu/kT}-1}, \quad u_\nu = \frac{8\pi\nu^2}{c^3}\left[\overline{U} - \frac{h\nu}{2}\right], \quad \overline{U} = \frac{h\nu}{e^{h\nu/kT}-1} + \frac{h\nu}{2}. \tag{9}$$

In the third equation of (9) we have also displayed the average energy $\overline{U}$ of a Planckian oscillator, containing the *zero–point energy* ½ $h\nu$ (Planck, 1911), which remains finite even at zero absolute temperature. It is interesting to note that, according to (7), the fractional part of the energy of the chaotic field, $\{E_\eta\}$, approaches exactly this value at high temperatures, $E_\zeta \to h\nu/2$. It can also be shown that the integer part $\xi \equiv [\eta]$ can be decomposed into a sum of independent binary random variables $\mathrm{u}_s = \mu_s(\xi)\cdot 2^s$ ($\mu_s = 0$ *or* 1), which satisfy Fermi–Dirac statistics (Varró, 2007), i.e.

$$\xi \equiv [\eta] = \sum_{s=0}^{\infty} \mathrm{u}_s = \sum_{s=0}^{\infty} \mu_s(\xi)\cdot 2^s,$$

$$P(\mu_s(\xi)=0) = \frac{1}{1+b^{2^s}}, \quad P(\mu_s(\xi)=1) = \frac{b^{2^s}}{1+b^{2^s}}, \quad b = \exp(-h\nu/kT) \quad (s = 0,\ 1,\ 2,...). \tag{10}$$

Equation (8) and (10) shows that there is a natural basis to express the possible photon excitation in terms of excitation of the '*binary photons*', represented by the fermionic variables $\mathrm{u}_s$. In the next section we shall show that a similar decomposition is possible for the fractional part $\zeta \equiv \{\eta\}$, too.

## 3. The statistics of the binary digits of the fractional part of the energy

In the present section we discuss the interrelation between the statistical properties of the fractional part of the scaled energy, $\zeta$ ($0 < \zeta < 1$), and the zeros and ones in its dyadic expansion. We express $\zeta$ as

$$\zeta = \sum_{k=1}^{\infty} \mathrm{v}_k = \sum_{k=1}^{\infty} \varepsilon_k(\zeta)/2^k, \quad \mathrm{v}_k \equiv \varepsilon_k(\zeta)/2^k, \quad \varepsilon_k(\zeta) = 0 \ \ or \ \ 1, \tag{11}$$

i.e., a possible realization of $\zeta$ may look like $z = 0.011010010...$ (with $\varepsilon_1 = 0,\ \varepsilon_2 = 1,\ \varepsilon_3 = 1,...$). Following Chatterji (1963), the algebraic construction of this problem can be formulated as follows. First, let $[\Omega_k, \mathscr{E}_k, P_k]$ ($k = 1,2,3,...$) be a sequence of measure spaces which defined as

$$\Omega_k = \{0,1\}, \quad \mathscr{E}_k = \{O, \Omega_k, \{0\}, \{1\}\}, \quad P_k(\{1\}) = p_k, \quad P_k(\{0\}) = 1 - p_k = q_k \ (0 \le p_k \le 1). \tag{12}$$

In the $\sigma-$algebra $\mathscr{E}_k$ the elementary sets {1} and {0} represent that in the physical system under discussion there appears an excitation of energy $h\nu/2^k$ (with probability $P(\{1\}) = p_k$) or not (with probability $q_k = P(\{0\})$), respectively. Then we construct the Cartesian product $[\Omega, \mathscr{E}, P]$ of the measure spaces $[\Omega_k, \mathscr{E}_k, P_k]$, i.e.

$$\Omega = \mathop{\times}_{k=1}^{\infty} \Omega_k, \quad \mathscr{E} = \mathop{\times}_{k=1}^{\infty} \mathscr{E}_k, \quad P = \mathop{\times}_{k=1}^{\infty} P_k, \tag{13}$$

and introduce $[I, \mathscr{B}, m]$, a measure space in the closed unit interval, where $I = [0,\ 1]$, the algebra $\mathscr{B}$ is built up from the Borel sets, and $m(A)$ denotes the Lebesgue measure of $A \in \mathscr{B}$. The expansion introduced in equation (11) corresponds to the mapping $\phi(\varepsilon) = \sum_{k=1}^{\infty} \varepsilon_k / 2^k$ of $\Omega$ onto $I$, where $\varepsilon = (\varepsilon_1, \varepsilon_2, ...)$ and the $\varepsilon_k$'s are 0's or 1's. According to this construction, the interrelation between the distributions of $\zeta$ and the $\varepsilon_k$'s is the interrelation of the probability measure $P$ on the product space $(\Omega, \mathscr{E})$ and the Lebesgue measure $m$ on the Borel space $(I, \mathscr{B})$. Besides various results, Chatterji (1963) has shown that the $\varepsilon_k$'s are independent, with respect to the absolutely continuous induced measure $m$, then the *density function of* $m$ have the form

$$f_a(x) = \begin{cases} 1 \ \ if \ \ a = 0 \\ \dfrac{a e^{ax}}{e^a - 1} \ \ if \ \ a \ne 0 \end{cases}, \quad f_{a=-\beta}(z) = f_\zeta(z) = \frac{\beta e^{-\beta \cdot z}}{1 - e^{-\beta}}, \quad a = -\beta = -h\nu/kT, \tag{14}$$

where $a$ is an arbitrary real number. It should be remarked that the measures with exponential densities are the only absolutely continuous measures for which the $\varepsilon_k$'s are independent. In our case, we have to put in (14) $a = -\beta = -h\nu/kT$, and then recover the density function $f_\zeta(z) = \beta e^{-\beta \cdot z}/(1 - e^{-\beta})$, which has been shown already in equation (4), as a result of our earlier work (Varró, 2007).

Before we prove the independence of the $\varepsilon_k$'s, we make some remarks on $\varepsilon_k(x)$'s considered as ordinary functions.

In the interval $(0,\ 1)$ the functions $\varepsilon_k(x)$ are piece-wise constant, namely

$$\varepsilon_k(x) = 0,\ 1,\ 0,\ 1,\ 0,\ ... \text{ on the intervals } \frac{q}{2^k} \le x < \frac{q+1}{2^k} \text{ with } q = 0,\ 1,\ ...\ 2^k - 1. \tag{15}$$

Accordingly, the function $\varphi_k(x) \equiv 1 - 2\varepsilon_k(x)$ will take on the values $+1,\ -1,\ +1,\ -1,\ +1,\ ...$ on the intervals shown in (15). Thus, the $\varepsilon_k(x)$'s are closely related to the Rademacher (1922) sytem of orthogonal functions { $r_0(x) = 1$, $r_k(x) = sign(\sin 2^k \pi x)$ $(k = 1,\ 2,\ ...)$}, in fact $\varphi_k(x) = r_k(x)$ for $k \ne 0$.

In order to prove the independence of the $\varepsilon_k(x)$'s in the case $a \ne 0$, by analogy from equation (10) for the distribution of the binary photons, we take

$$P_a(\varepsilon_k(x)=1)=\frac{e^{a/2^k}}{1+e^{a/2^k}}\,,\qquad P_a(\varepsilon_k(x)=0)=\frac{1}{1+e^{a/2^k}}\,,\qquad P_a(\varepsilon_k(x)=\delta_k)=\frac{e^{a\delta_k/2^k}}{1+e^{a/2^k}}\,,\tag{16}$$

where $\delta_k=1$ or $\delta_k=0$. The products of such probabilities reads

$$\prod_{k=1}^{n}P_a(\varepsilon_k(x)=\delta_k)=\frac{\exp(a\sum_{k=1}^{n}\delta_k/2^k)}{\prod_{k=1}^{n}(1+e^{a/2^k})},\qquad \delta_k=1 \text{ or } \delta_k=0\,.\tag{17}$$

Since $(1-x)(1+x)(1+x^2)\cdot\ldots\cdot(1+x^{2^s})=1-x^{2^{s+1}}$, by taking $x=(e^{a/2^k})$ we have

$$\frac{1-(e^{a/2^k})^{2^k}}{1-(e^{a/2^k})}=(1+(e^{a/2^k}))\cdot(1+(e^{a/2^k})^2)\cdot(1+(e^{a/2^k})^2)\cdots(1+(e^{a/2^k})^{2^{k-1}})\,,$$

$$\frac{1}{\prod_{k=1}^{n}(1+e^{a/2^k})}=\frac{e^{a/2^n}-1}{e^a-1}\,,\tag{18}$$

thus the product of the probabilities can be brought to the closed form

$$\prod_{k=1}^{n}P_a(\varepsilon_k(x)=\delta_k)=\frac{[\exp(a\sum_{k=1}^{n}\delta_k/2^k)](e^{a/2^n}-1)}{e^a-1}\,.\tag{19}$$

On the other hand, because $x=\sum_{k=1}^{\infty}\varepsilon_k(x)/2^k$, the relation $\sum_{k=1}^{n}\delta_k/2^k\le x<\sum_{k=1}^{n}\delta_k/2^k+1/2^n$ means the joint event $\{\varepsilon_1(x)=\delta_1\}\cdot\{\varepsilon_2(x)=\delta_2\}\cdots\{\varepsilon_n(x)=\delta_n\}$, and since the density function (14) is in fact the derivative $f_a(x)=\dfrac{ae^{ax}}{e^a-1}=\dfrac{1}{e^a-1}\dfrac{d}{dx}e^{ax}$ ,

we may write

$$P(\varepsilon_k(x)=\delta_k;1\le k\le n)=\int_{\sum_{k=1}^{n}\delta_k/2^k}^{\sum_{k=1}^{n}\delta_k/2^k+1/2^n}f_a(x)dx=\frac{1}{e^a-1}\left[\exp(a\textstyle\sum_{k=1}^{n}\delta_k/2^k)\right](e^{a/2^n}-1)\,.\tag{20}$$

By comparing (20) and (19) we see that the joint probability (20) of the products of events $\{\varepsilon_k(x)=\delta_k\}$ equals the product (19) of the probabilities of these events, i.e.

$$P(\varepsilon_1(x)=\delta_1,\varepsilon_2(x)=\delta_2,\cdots,\varepsilon_n(x)=\delta_n)=P(\varepsilon_1(x)=\delta_1)P(\varepsilon_2(x)=\delta_2)\cdots P(\varepsilon_n(x)=\delta_n)\,.\tag{21}$$

Equation (21) shows the independence of $\varepsilon_k(x)$-s with respect to the measure whose density is $f_a(x)$, which coincides our $f_\zeta(z)$ in (4), by choosing $a=-\beta=-h\nu/kT$ . On the basis of (10) and (16) our results can be summarized as follows

$$\eta=\xi+\zeta\,,\quad \xi=\sum_{s=0}^{\infty}\mathrm{u}_s=\sum_{s=0}^{\infty}\mu_s(\xi)\cdot 2^s\,,\quad \zeta=\sum_{k=1}^{\infty}\mathrm{v}_k=\sum_{k=1}^{\infty}\varepsilon_k(\zeta)\cdot 2^{-k}\,,\tag{22}$$

or, by introducing a uniform notation $\lambda_r$ for $\mu_r$ and $\varepsilon_r$

$$\eta=\sum_{r=-\infty}^{\infty}\lambda_r\cdot 2^r\,,\quad \lambda_r=\mu_r(\xi),\ r\ge 0 \text{ and } \lambda_r=\varepsilon_{-r}(\zeta),\ r<0\,.\tag{23}$$

Equations (22) and (23) show that the energy of the chaotic radiation can be decomposed to an integer part and to a fractional part. The distribution of the integer part is just the Planck – Bose distribution (8), which can be further decomposed to the Fermi – Dirac distributions (10) of binary photons (Varró, 2007). The fractional part of $\eta=[\eta]+\{\eta\}\equiv\xi+\zeta$ has been expanded in the dyadic representation (11), whose factors, $\varepsilon_k(x)$, have been turned out to be independent random variables.

In the first case $(a=0)$ shown in equation (14) the distribution function is unity, $f_{a=0}(x)=1$ and the expectation value of the fractional part of the energy would be $\bar{\zeta}_0=\int_0^1 f_{a=0}(x)xdx=1/2$, which yields just the *zero–point energy,* $h\nu\cdot\bar{\zeta}_0=h\nu/2$. However, this expectation $\bar{\zeta}_0$ of a scaled energy $\zeta_0$ cannot stem from the original chaotic distribution of $\eta$, because for the fractional part of $\eta$, for $\zeta=\{\eta\}$, we have already derived the temperature dependent result $f_{a=-\beta}(z)$, displayed in equations (14) and (4). One may associate the random fractional energy $\zeta_0$ to the system which is being in thermal equilibrium with the black–body radiation. This was first done by Planck (1911) himself, who discovered the zero–point energy, as a 'by-product' of his second theory. As is shown in (9), the average mode energy is the difference of the oscillator's energy and the zero–point energy of the oscillator, i.e. $\bar{U}-h\nu/2$. Owing to the *same distributions* of $\varepsilon_k(\zeta_0)$-s, this 'zero–point variable' $\zeta_0$ may be considered as a physical representative of an *ideal random number generator*.The further study of the possible physical meanings of the distributions of $\varepsilon_k(x)$-s in this special case, is beyond the scope of the present paper. We leave this discussion to a separate paper (Varró, 2012).

## 4. Summary

In Section 2 we summarized our earlier results on the fractional part, $\zeta=\{\eta\}$, and on the integer part, $\xi=[\eta]$, of the scaled energy $\eta=E/h\nu$ of a mode of black–body radiation, which has been derived from a two-dimensional Gaussian distribution the amplitudes. The integer part satisfies the Planck–Bose distribution, which can be decomposed to Fermi distributions of binary photons. In Section 3 we have represented the fractional part by binary sequences $\{\varepsilon_k(\zeta)\}$, and shown the algebraic structure behind the interrelations of the distribution of $\zeta$ and the distributions of the digits $\{\varepsilon_k\}$ which represent it. We have also shown the close connection of the functions $\varepsilon_k(x)$ with Rademacher's orthogonal system. We have given the distribution of these binary digits, and proved the independence of them. According to an earlier mathematical theorem, we have shown that, if the 'digital randomness' is such that the binary variables at different positions are independent, then we recover the distribution of the fractional part of the energy of the thermal radiation mode, which coincides with our earlier result. If, in addition, 'the randomness is maximal', i.e. when these digits have the same distribution, then the associated new variable $\zeta_0$ has a uniform distribution on the interval $[0,1]$, and its expectation value yields just the zero–point energy, $h\nu\cdot\bar{\zeta}_0=h\nu/2$. According to the present description, this cannot be associated to the photon mode, but rather, it can be a property of a resonant absorber, like the Planckian oscillators being in thermal equilibrium with the mode. We think that the above considerations and results may have even technological relevance in the contex of thermal noise and physical round-off errors.

**Acknowledgments**
This work has been supported by the Hungarian National Scientific Research Foundation OTKA, Grant No. K 104260. Partial support by the National Development Agency, Grant No. ELI_ 09-1-2010-0010, Helios Project is also acknowledged.